\documentclass[12pt]{article} %hier wegkommentieren

\usepackage[ngerman,english]{babel}
\useshorthands{"}
\addto\extrasenglish{\languageshorthands{ngerman}}
\usepackage[utf8]{inputenc}
\usepackage[T1]{fontenc}
\usepackage{amsmath}
\usepackage{graphicx}
\usepackage{amssymb}
\usepackage{hyperref}
\usepackage{amsmath}
\usepackage{mathrsfs}
\usepackage{color}
\usepackage{subfig}
\usepackage{amsthm}

\theoremstyle{definition}

\renewcommand{\theta}{\vartheta}

\title{{\Large\bf Microscopic Origin of Page Curve}}
\author{{\bf Artem Averin$^{\textrm{a}}$\footnote{artem.averin@campus.lmu.de}}}

\begin{document}

\maketitle

%\vskip-2.5cm
\centerline{\it $^{\textrm{a}}$ Arnold--Sommerfeld--Center for Theoretical Physics,}
\centerline{\it Ludwig--Maximilians--Universit\"at, 80333 M\"unchen, Germany}
%\medskip
%\centerline{\it $^{\textrm{c}}$ Center for Cosmology and Particle Physics,
%Department of Physics, New York University}
%\centerline{\it 4 Washington Place, New York, NY 10003, USA}
%\medskip
%\centerline{\it $^{\textrm{d}}$ Instituto de F\'{\i}sica Te\'orica UAM-CSIC, C-XVI,
%Universidad Aut\'onoma de Madrid,}
%\centerline{\it Cantoblanco, 28049 Madrid, Spain}

\vskip1cm
\begin{abstract}
{{
Applications of the Ryu-Takayanagi formula to evaporating black holes are known to reproduce the Page curve, although leaving open the microscopic origin of the necessary black hole degrees of freedom responsible for the entanglement and microstates. Here, we fill this gap by utilizing a recently proven generalized Ryu-Takayanagi formula which keeps the contact to the theory's Hamiltonian phase space. Precisely, we show that in any diffeomorphism invariant field theory containing stationary black holes with bifurcate Killing horizons the phase space states distinguishable by their Hamiltonian surface charges over the bifurcation surface must provide the degrees of freedom responsible for the black hole's Wald entropy. This result is part of a larger program where measurable quantities of a quantum theory are expressed as weighted sums over paths in the theory's phase space. Suited reorganization tools for those sums make then physical phenomena evident which might be obscured in a naive summation relying on spacetime notions such as field configuration spaces and Feynman diagrams. Besides presenting this result, we provide here a conceptual overview of the program as an entry point for unfamiliar readers explaining the central notion of the program and why it reveals the results obtained so far. These include especially the gravitational entropy bound and the here applied generalized Ryu-Takayanagi prescription shown earlier within the program. Using the information paradox as a leitmotif, our proposed program uncovers black hole hair that may be hidden within a naive spacetime interpretation.          
}}
\end{abstract}

%\begin{flushright}
%\vskipcm
%{\small  MPP--2016--3},
%{\small LMU--ASC 05/16}
%\end{flushright}

\newpage

\setcounter{tocdepth}{2}
\tableofcontents
\break

\section{Manifesto on the End of Naive Summation}
\label{Kapitel 1}

A spectre is haunting the current understanding of the fundamental laws of Nature - the spectre of the necessity of spacetime. 

The exorcism of this spectre, as we demonstrate in this overview, is indispensable for exposing the true alliance among the phenomena governed by the fundamental laws. 
\\

Such a phenomenon is the existence of black holes. Obviously, this is a spacetime phenomenon and despite their direct observation, there are still open problems concerning black holes. Most famously, the black hole information paradox is still considered to be not resolved although some progress has been made recently (see \cite{Almheiri:2020cfm} for a review of the recent developments including a review of the information paradox itself).

In essence, a formula for the entanglement entropy in gravitational theories was developed which is motivated by the analogous Ryu-Takayanagi formula for entanglement entropy in AdS/CFT. When this newly developed formula is applied to infer the time dependence of the entanglement entropy of an evaporating black hole, the Page curve is directly recovered within the gravitational theory. This Page curve is what is required by unitarity of quantum mechanical evolution.

However, while these new developments show that the knowledge about quantum mechanical unitarity is somehow incorporated in the gravitational theories, they do not make a statement about precisely how. More explicitly, the obtained late-time behavior of the Page curve means that the black hole should possess degrees of freedom responsible for the associated entanglement entropy. What are those black hole degrees of freedom? Or, in other words, what is the microscopic origin of the Page curve? This is the same open question that arises when considering the Bekenstein-Hawking entropy of a black hole. Its many derivations, as the one of Gibbons and Hawking \cite{Gibbons:1976ue}, show that black holes should possess degrees of freedom responsible for an entropy in accordance with the Bekenstein-Hawking formula. However, these arguments leave open the question about what those black hole degrees of freedom are. What is the microscopic origin of the black hole entropy?

This is related to the original formulation of Hawking's information paradox. If a black hole forms in gravitational collapse, and eventually evaporates by Hawking radiation, how to get back the information about the original formation of the black hole? In ordinary black body radiation, detecting the corrections to thermality of the radiation allows, in principle, to obtain the information about the precise black body microstate. However, in order for the same to be true for a black hole, a black hole must possess enough microstates to capture the information about its formation. According to the mentioned Bekenstein-Hawking law, the gravitational theory suggests black holes to indeed possess microstates. However, so far there is no consensus on where these black hole degrees of freedom are in the gravitational theory.

Precisely this question takes the center stage in the discussion of the information paradox, especially after the recent developments concerning the recovery of the Page curve within the gravitational theory. Indeed, the recovery of the Page curve within the gravitational theory means that the black hole degrees of freedom are enough to capture the information about its formation. It then remains to find these black hole degrees of freedom responsible for black hole microstates and the entanglement entropy in the Page curve within the gravitational description. We hence identify this as the central question of the black hole information paradox.

Unfortunately, the formula for the gravitational entanglement entropy used in obtaining the Page curve (see the review \cite{Almheiri:2020cfm} and the original references \cite{Almheiri:2019psf,Penington:2019npb,Almheiri:2019hni,Penington:2019kki,Almheiri:2019qdq}) makes no statement about the microscopic origin of the degrees of freedom responsible for entanglement entropy. The situation is analogous to that of the Bekenstein-Hawking entropy of black holes. The gravitational theory predicts its existence but seems to lack a microscopic explanation of the involved degrees of freedom within the gravitational theory itself. 

Here, we want to precisely fill this gap in the literature. The new point we add to the discussion is that we will show that the gravitational theory in fact does make a statement about the microscopic origin of black hole degrees of freedom responsible for black hole entropy and entanglement entropy in the Page curve. We point out that a gravitational theory is even forced by diffeomorphism invariance to contain those black hole degrees of freedom. To make this point evident, however, we will replace, in a sense, the notion of spacetime by a more general concept. Note that replacing spacetime emerges as a dialectic necessity to unravel the physics behind the information paradox of black holes which are by themselves spacetime phenomena. 

Fortunately, to accomplish our aim, the technical work is already done in \cite{Averin:2024its,Averin:2024yeo,Averin:2025zua}. In fact, our main result concerning the microscopic origin of black hole degrees of freedom is an application of the generalized Ryu-Takayanagi formula formulated and proven in \cite{Averin:2025zua}. One important novelty of this generalized formula for entanglement entropies in gravitational theories is the contact to the theory's Hamiltonian phase space. It is this contact to phase space where it goes beyond the analogous formula used in the developments \cite{Almheiri:2019psf,Penington:2019npb,Almheiri:2019hni,Penington:2019kki,Almheiri:2019qdq} and which reveals the microscopic origin of involved black hole degrees of freedom. 

We will present this as our main result in chapter \ref{Kapitel 3}. Since it is a direct consequence of \cite{Averin:2024its,Averin:2024yeo,Averin:2025zua}, there will be no need for technical calculations here. Instead, we want to use the opportunity to give an overview of \cite{Averin:2024its,Averin:2024yeo,Averin:2025zua} which emphasizes the conceptual novelties in an intuitively understandable way. As is common in quantum field theory, the technicalities of \cite{Averin:2024its,Averin:2024yeo,Averin:2025zua} are quite abstract and involved although the concepts and physics are very vivid and intuitively accessible. In what follows, we provide an overview of those concepts and also give an entry point for readers not acquainted with \cite{Averin:2024its,Averin:2024yeo,Averin:2025zua}. We explain why the concepts introduced there provide a new understanding of certain physical phenomena including the information paradox which we eventually treat in chapter \ref{Kapitel 3}.
\\

We start our overview of \cite{Averin:2024its,Averin:2024yeo,Averin:2025zua} by considering the same and simple situation as in the beginning of \cite{Averin:2024its}. We let a coin fall down and are interested in measuring the value of a chosen observable $O$ in this experiment. The role we expect from the fundamental laws of Nature is to predict the measured value $O$ a priori. In their most modern form, they can be paraphrased as 

\begin{equation}
O = \sum_{\text{possibilities}} (\text{weight factors})
\label{1}
\end{equation} 

which means that the observable is determined by a weighted sum over all possibilities the system could evolve, i.e. we have to sum up certain weight factors that are associated to each possible way the coin could behave. The falling coin is here of course just an illustrative surrogate for the evolution of any physical system with \eqref{1} being still valid. What is important in the prescription \eqref{1}, is that a measurable quantity is determined as a weighted sum over all possibilities the system could evolve.

The prescription \eqref{1} provides the underlying structure of quantum theory which is expected to be the theoretical framework the fundamental laws of Nature should be formulated in. Indeed, \eqref{1} reflects the quantum mechanical superposition principle, the fundamental property of quantum mechanics, and can be derived from the Schrödinger equation. A derivation of \eqref{1} is given in \cite{Averin:2024its} where also the technicalities involved in the formalized version are introduced and discussed in detail. Since we focus here merely on the concepts, the qualitative version \eqref{1} will be sufficient for the present discussion.

The pair $(\text{possibilities}, \text{ weight factors})$ of a set of possibilities with associated weight factors constitute the definition of the particular theory under consideration used for the description of a concrete physical system. Their precise mathematical meaning is given as mentioned in \cite{Averin:2024its}. 

At this point, one might ask whether quantum mechanics (or, equivalently, the prescription for observables \eqref{1}) is enough for the description of the real world? Could there be anything beyond be of need? There are good reasons to believe that quantum mechanics provides the appropriate framework to describe the fundamental laws. After all, within the standard model of particle physics, its predictions have matched with experiments extremely well. Gravity is not contained in the standard model. However, according to the AdS/CFT correspondence, a certain quantum theory of gravity is dual to a certain ordinary quantum theory without gravity. This means that quantum theory should also in principle be the right framework in the description of gravitational phenomena. A priori, there seems to be no need for extending quantum mechanics. Then, however, we must in principle be able to explain all of Nature's phenomena, in particular those of quantum gravity, by \eqref{1}. This is precisely the viewpoint we want to advocate here. Our program can be summarized as the aim to explain phenomena by \eqref{1} once the summation is organized properly. This is also for the skeptic of quantum mechanics a necessary program since he needs to prove its failure.

Having set the framework, we can immediately note a remarkable consequence. The fundamental laws of Nature make according to \eqref{1} only statements about the functional relations of measurable quantities but they cannot say anything about how the set $(\text{possibilities}, \text{ weight factors})$ is materialized in Nature, i.e. its substance. This is because an observer has only access to the measurable quantities. We can think of them as being the observer's window into the real world while he uses \eqref{1} together with a mathematical construction of the set $(\text{possibilities}, \text{ weight factors})$ in his inner room of mind to understand the relations among measurable quantities.

This thought fixes then the role of a physicist. In principle, it is the construction of a set $(\text{possibilities}, \text{ weight factors})$ which is appropriate for a description of Nature.

This leads to the question of what the right set (possibilities, weight factors) is? The quest of finding it has a long history culminating in the 1970s with the formulation of the standard model of particle physics. Herein, the pair $(\text{possibilities}, \text{ weight factors})$ is realized by a set of different fields interacting on spacetime and the corresponding quantum theory, i.e. the prescription \eqref{1}, is adequate for an accurate description of the phenomena of particle physics observed so far and where the summation in \eqref{1} can be evaluated. Despite its success, the standard model has some shortcomings. These are for instance phenomena which obviously require its extension such as dark matter or dark energy. However, there are also observed phenomena which are expected to be explainable by the standard model but a precise description is lacking because no way of evaluating the required summations in \eqref{1} is known. An example of such a phenomenon is color confinement. Lattice computations or AdS/CFT arguments suggest that quantum chromodynamics explains this observed phenomenon in principle. However, a full theoretical description would require the evaluation of perturbative expansions in a strong coupling regime. 

We are here precisely interested in those types of phenomena. In fact, our main point is to stress that the conventional way of organizing the sum of possibilities in \eqref{1} may possess some shortcomings hiding the explanation of physical phenomena. We point out that the origin of at least some physical phenomena becomes directly evident once the sum in \eqref{1} is properly organized.

Before we can make our point and explain how to smartly order \eqref{1}, we should discuss what the conventional way is and why it may lead to shortcomings which we should overcome. 

To illustrate the conventional summation, we take a look at the standard model. It is a theory of fields interacting on spacetime. And as in any quantum field theory, the sum of possibilities \eqref{1} is performed by summing over all processes which can happen at any point in spacetime. Technically, this is given by a perturbative expansion in terms of Feynman diagrams. Note that the central concept determining the organization of the sum of possibilities is that of spacetime.

This way of organizing the sum of possibilities \eqref{1} as a sum of all possible processes happening at all possible points in spacetime is natural for a field theory. It leads to a perturbative expansion and hence to a practical way of evaluation of the sum \eqref{1}. Predictions made this way within the standard model are in excellent agreement with experiments. However, the crucial ingredient here is that the obtained perturbative expansion is reliable and computable. Whether this is the case depends on the concrete observable being computed, i.e. on the question being asked. Spacetime being the central paradigm in organizing \eqref{1} may be appropriate for some questions and for some other not. In the latter case, the organization of the summation of the possibilities then should, if possible, be replaced by something else. We illustrate the need for such reorganizations with a very simple analogy. A common question in quantum field theory is the sensitivity on a particular length scale, i.e. how sensitive is a specific observable on processes happening at a certain length scale? As a simplified analogy to that question, we consider the alternating harmonic series

\begin{equation}
\sum_{k=1}^{\Lambda} \frac{(-1)^{k+1}}{k}
\label{2}
\end{equation}

and ask for the sensitivity on the cutoff $\Lambda$? Looking at the terms with odd denominator suggests a logarithmic dependence on $\Lambda.$ The terms with even denominator lead to the same suggestion. However, following the the naive suggestion of \eqref{2} depending logarithmically on the cutoff $\Lambda$ is wrong. In fact, this naive conclusion does not recognize the delicate cancellations between the terms with even and odd denominators in \eqref{2} leading to an insensitivity to $\Lambda,$ i.e. the convergence of the series in \eqref{2} for $\Lambda \to \infty.$ This simple example illustrates that physical phenomena may be hidden within a naive organization of the quantum mechanical sum of possibilities \eqref{1}. Of course, the sum \eqref{1} is commutative and does not depend on the way of organizing it. However, on the one hand, a physical phenomenon might not be directly evident in a Feynman diagram expansion with spacetime being the paradigm of ordering. On the other hand, the same phenomenon could reveal itself if a different and more appropriate ordering is used. 

Our considerations so far show us the necessity to study different ways of organizing the quantum mechanical summation \eqref{1}. How do we find sensible ways of organizing the sum \eqref{1} in a useful way? This is the central question in our approach we are advocating here and it directly leads us to a new concept. We can motivate an answer to this question by considering and generalizing the way the conventional Feynman diagram expansion is organized in quantum field theory. As explained, the concept of spacetime determines the organization. However, in quantum field theory, one is often asking questions concerning observables accessible above a certain length scale (or, equivalently, below a certain energy scale). In the summation \eqref{1}, one then partitions the low-energy modes of quantum fields involved in processes happening at low energy into bundles while the high-energy modes are integrated out. This leads to the renormalization group flow. The bundle of low-energy processes is governed by the Wilsonian effective action involving running couplings. 

The renormalization group flow is an extremely powerful tool in quantum field theory. Couplings running with the energy scale reveal the origin of asymptotic freedom. This, in turn, was central in identifying non-Ablelian gauge theories as governing the elementary particle interactions. 

We observe that the line of thought leading to the renormalization group flow is generalizable. The idea is to obtain further powerful tools by exploiting this generalization which may reveal other aspects of the structure of given quantum theories. 

In order to do so, we reconsider the line of thought leading to the renormalization group flow while we this time emphasize the aspects allowing for a generalization. The starting point of the argumentation is again the sum of possibilities \eqref{1} which is used to compute a measurable quantity of a given theory defined by a set $(\text{possibilities}, \text{ weight factors}).$ We can then partition the sum \eqref{1} into suited bundles of possibilities with the aim of this suited bundling to make physical phenomena evident which are otherwise hidden in a naive summation.

Contrary to the previous argumentation, the bundles do not need to consists of the low-energy modes of quantum fields which is a concept relying on the notion of spacetime. Instead, it is natural to allow the bundles to be a suited subset of the set $(\text{possibilities}, \text{ weight factors}).$ The bundles are hence submanifolds of the set of all possibilities and were termed possifolds in \cite{Averin:2024its} where this concept was introduced.

While the possifold notion appears to be a natural concept, we so far have not explained what a suited bundle of possibilities is. When is it considered suited? What makes it better than the bundle of low-energy modes of the previous argument? The answer is that the suitability depends on the questions being asked. In the previous argument the bundle of low-energy modes was appropriate since the asked questions were concerning observables accessible at low energies or large length scales, respectively. For other types of questions such a bundling may not be useful, especially when the questions do not obviously relate to spacetime concepts such as low or high energies.

What are examples for such questions? They are easily found in black hole physics. Suppose, we are considering a large black hole and are asking questions concerning its internal structure. That is, we are measuring observable quantities and a quantum theory aimed at describing the black hole must be given, as explained, by a set $(\text{possibilities}, \text{ weight factors})$ which can via \eqref{1} reproduce correctly our measurements. Large black holes are expected to obey the laws of thermodynamics and as such have an entropy and an associated large number of microstates. With the phrase asking questions concerning the internal structure, we mean questions involving those black hole microstates. For instance, we could measure transition amplitudes among microstates or simply their degeneracy (which should reproduce the black hole entropy). Then, what is the appropriate possifold for this type of questions? Put it differently, what is the relevant subset of the set of all possibilities $(\text{possibilities}, \text{ weight factors})$ which are relevant in the summation \eqref{1} for reproducing observables concerning black hole microstates, e.g. black hole entropy? They are not obviously related to the low-energy or high-energy modes of the quantum fields present in the given theory. This conclusion is reached because on the one hand large black holes should be governed by low-energy processes. On the other hand, large black holes can be produced in high-energy collisions. We hence learn that the bundle of possibilities needed to understand black hole physics goes beyond the common spacetime concepts of separating small and large distance physics. Furthermore, as we have explained, the proper bundle of possibilities must contain the information about the black hole microstates and is therefore central in resolving the information paradox. 

In summary, the arguments show how black hole physics and the information paradox enforce the introduction of the possifold concept.

Then, one is immediately driven to the question on what the concrete bundle of possibilities suited for the description of black hole microstates is? How to describe it in a given theory $(\text{possibilities}, \text{ weight factors})$ concretely which contains black holes (such as, for instance, Einstein gravity)? And does the theory contain the required bundle of possibilities for the description of black holes or should it be added by hand?

Indeed, we will show in chapter \ref{Kapitel 3} that diffeomorphism invariant field theories containing black holes do accommodate the mentioned required bundle of possibilities and that its specific form is highly restricted.

This result means that, as already explained, gravitational theories contain the necessary degrees of freedom to account for black hole microstates with the corresponding implications for the information paradox. We will obtain this result by an application of the generalized Ryu-Takayanagi formula formulated and proven in \cite{Averin:2025zua}.

This generalized Ryu-Takayanagi formula is part of a series of results which can be formulated and proven using the possifold concept whose advocation is the main purpose of this chapter. As explained, the possifold concept can be used in the explanation of some physical phenomena in a recurring way. By partitioning the quantum mechanical sum of possibilities \eqref{1} into suited bundles of possibilities (possifolds), some physical phenomena may become evident which are otherwise hidden in a naive summation.

The formalism, language and notation for mathematically carrying out this idea was developed within the framework introduced in \cite{Averin:2024its}. It was then used in \cite{Averin:2024yeo} to formulate and prove the gravitational entropy bound for arbitrary diffeomorphism invariant field theories. As an implication of this derivation, the mentioned generalized Ryu-Takayanagi formula was shown in \cite{Averin:2025zua}.

In this chapter, we have argued for the necessity of the possifold concept for an understanding of the fundamental laws of Nature. We pointed out that naive summation of the quantum mechanical sum of possibilities such as relying on the spacetime concepts fails in revealing the origin of certain phenomena. The results listed in the previous paragraph demonstrate indeed that the possifold concept provides the physical and mathematical origin of phenomena which previously only were conjectured heuristically or formulated in the context of AdS/CFT.

Before applying the possifold concept to the information paradox in chapter \ref{Kapitel 3}, we note that the concept is only powerful if the bundles of possibilities are grouped in a suited way. We already explained that the meaning of what is suited depends on the questions being asked. Possifolds and the asking of questions are therefore tightly connected. In the next chapter, we therefore focus on the aspect of asking questions. 

\section{Question of Swampland or Swampland of Questions?}
\label{Kapitel 2}

We have explained in the last chapter in a qualitative manner what the possifold concept is and why its need emerges naturally. We have seen possifolds within a given quantum theory to be related to the measurable quantities one is interested in, i.e. to the asked questions. In this chapter, we will further investigate this relation. We start with the concept of asking questions within a quantum theory. Although this concept might appear obvious at first sight, we will emphasize some remarkable aspects which may cause (and in some cases indeed previously caused) considerable confusion.

We start by considering a quantum theory given by a set (possibilities, weight factors). This theory is to describe given physical phenomena such as the falling coin of the last chapter or the entire fundamental laws of Nature. Suppose, we are now asking a question about a considered phenomenon. What does this concretely mean? It means we want to use our given theory to predict some measurable quantity. We therefore have to map our question to the concrete measurable quantity it is asking for which we then evaluate according to the summation \eqref{1} within the considered theory. 

This procedure can be not as obvious as it sounds. For instance, in quantum field theory one is often interested in predicting differential cross sections to be measured in particle scattering experiments. The former are inferred from scattering amplitudes which in turn are read off from the Laurent coefficients of momentum-space correlation functions. The latter are the quantities given by a quantum mechanical summation of the form \eqref{1}. This example shows that the mapping of the concrete question to the observable to be computed via \eqref{1} can be quite abstract. Furthermore, it is not assured that a given question can be mapped onto a sensible observable.

Note that for a quantum theory given by a set (possibilities, weight factors), all measurable quantities are determined by the relation \eqref{1}. In other words, specifying a concrete theory fixes the set of measurable quantities one could in principle observe. Put it differently, it fixes the list of all sensible questions one can ask within that theory. A question that cannot be mapped to measurable quantities of the theory is not on the list and simply makes no sense to be asked. This observation is both remarkable and trivial. We will give simple and less simple examples of not allowed questions in a moment.
\\

To give an example for a not allowed question, we go back to our considered theory $(\text{possibilities}, \text{ weight factors}).$ The weight factors determining the contribution of a particular possibility in the quantum mechanical summation \eqref{1} depend on Planck's constant $\hbar.$ In the classical limit $\hbar \to 0,$ there are typically dominant possibilities determining the value of a measurable quantity in \eqref{1}. Observing a physical system in this limit, i.e. measuring observable quantities, it looks \emph{as if} the evolution of the system follows a definite possibility. For instance, an observer of a falling coin would construct the fiction or the illusion that the system evolves \emph{as if} following a definite trajectory. This limit is the realm of validity of classical physics whose constructed illusion is reflected in the experiences of daily life. As a consequence, one is naively often driven to ask questions which are artifacts of this illusion. For finite $\hbar,$ however, such questions need not to make sense. For instance, for the falling coin, we can ask for its position and momentum at an instant of time. While this question totally makes sense in the illusion of definite trajectories governed by classical equations of motion, this question makes no sense for finite $\hbar.$ It provides an example of a not allowed question.
\\

This example teaches us that there are questions which do make sense in the constructed illusion associated to a quantum theory but are actually not allowed in the quantum theory itself. The example shown above provides a very simple case of this scenario. However, we will see below more complicated cases which were the source of a lot of confusion. In fact, one quite often encounters questions which are artifacts of the constructed illusion but are actually not allowed in a given quantum theory. This should not surprise. The constructed illusion is not only a source of motivation for the questions to be asked in a quantum theory but for the set $(\text{possibilities}, \text{ weight factors})$ defining the theory itself.

So far, we have taken as given the defining set (possibilities, weight factors) of a theory which is supposed to describe physical phenomena of interest. But how is this set obtained for concretely given phenomena? Finally, the goal is to give a theory $(\text{possibilities}, \text{ weight factors})$ which accurately describes the fundamental laws of Nature. But how can this set be obtained? The guiding principle is the classical limit, the constructed illusion as experienced in daily life. The requirement on a set $(\text{possibilities}, \text{ weight factors})$ to reproduce the constructed illusion puts severe restrictions on that set. It permits to conclude the set $(\text{possibilities}, \text{ weight factors})$ from the behavior of a physical phenomenon in the classical limit. This is the process of quantization.

In the process of quantization, the quantum theory (possibilities, weight factors) is obtained from the constructed illusion. During this process, however, sensible questions of the illusion may become senseless in the quantum theory as we have already seen. Already to find the sensible questions of a given quantum theory $(\text{possibilities}, \text{ weight factors}),$ it therefore appears to be of central importance to study the mathematical structure of the set $(\text{possibilities}, \text{ weight factors}).$ We have seen the questions of a quantum theory $(\text{possibilities}, \text{ weight factors})$ to be related to its possifolds which in turn may reveal its physical phenomena.

In summary, our arguments show the emergence of a necessary approach for the understanding of a quantum theory $(\text{possibilities}, \text{ weight factors})$ in which the possifold notion is central. The idea is to study the quantum mechanical sum of possibilities \eqref{1} by partitioning using suited possifolds. The aim of this manipulation is then to trace back information about the mathematical structure $(\text{possibilities}, \text{ weight factors})$ which encodes the physics of the considered quantum theory. 

We have now conceptually motivated and outlined our advocated program to study quantum theory with the necessary use of possifolds. Quite at this point, the question is whether it is useful? Does the explained possifold program lead to non-trivial results? And is the possifold program able to reveal results which cannot be reached within conventional methods?

Of course, we cannot give an exhausting answer to these questions. However, in the reminder of this chapter, we first mention results which have been obtained within the possifold program. Next, as an outlook, we argue where the possifold program should furthermore reveal important insights. 
\\

To briefly explain the results the possifold program achieved so far, we have to go slightly more into the technical details. The quantum mechanical sum of possibilities \eqref{1} is concretely speaking a weighted sum over paths in the Hamiltonian phase space of the considered theory. 

For diffeomorphism invariant field theories, the possifold program then leads to severe bounds on the phase space volumes of submanifolds in the theory's phase space. Indeed, these bounds were shown in \cite{Averin:2024yeo} to emerge by reorganizing the sum over paths in phase space while exploiting diffeomorphism invariance as suggested by the advocated approach of the possifold program.

Why is this result important? First, it provides a concrete example how the possifold program can be used to obtain non-trivial statements. Note that the approach can from a purely mathematical perspective be seen as a tool to study the geometry of phase space by considering suited weighted sums over paths in phase space. The precise form of those sums is hereby motivated quantum mechanically. A suited reorganization of this sum makes hidden properties of the phase space geometry evident. The example illustrates how the mathematical structure of the set $(\text{possibilities}, \text{ weight factors})$ defining a quantum theory (which is here the phase space geometry) is made evident by considering and reorganizing the sum of possibilities \eqref{1} into suited bundles. The mathematical structure of the set $(\text{possibilities}, \text{ weight factors})$ then might reflect a non-trivial property of the considered theory. 

This is precisely the case for the bounds on phase space volumes obtained in \cite{Averin:2024yeo} for general diffeomorphism invariant field theories. They provide a precise formulation of what on a heuristic basis is conjectured to be the gravitational entropy bound in a quantum theory of gravity. The heuristic motivation for the gravitational entropy bound in turn comes from the laws of black hole thermodynamics (see \cite{Averin:2024yeo} and references therein). Remarkably, the possifold program, here in particular the interplay of the commutativity of the quantum mechanical sum \eqref{1} together with diffeomorphism invariance, reveals the origin of the gravitational entropy bound. 

Besides being able to explain its origin, the possifold program also shows that the gravitational entropy bound in a quantum theory of gravity is not a choice. Rather, it is an unavoidable consequence of the interplay of quantum mechanics (commutativity of the sum of possibilities \eqref{1}) together with gravity (diffeomorphism invariance). Although not much is known about the behavior of quantum gravity, the laws of black hole thermodynamics are among the most robust predictions. They are often used (maybe on a heuristic basis) to obtain or at least conjecture generic statements any quantum theory of gravity should obey. Arguments for such conjectures often rely on the mentioned heuristic versions of the gravitational entropy bound. The idea then is that there is a set of theories which fails to satisfy these required statements. This set of theories is then said to be in the swampland (see \cite{Palti:2019pca} for a review of this approach). Although there are several sources for those conjectured statements being able to rule out theories possessing no consistent coupling to gravity, which are then called swampland conjectures, the ones relying on black hole physics are believed to be quite robust. In any case, the possifold program demonstrates that there cannot be any diffeomorphism invariant field theory which fails to respect the gravitational entropy bound (or any swampland conjecture relying solely on it). It is not a choice or anything that has to be put in by hand to ensure consistency. Contrary, it turns out as a geometrical phase space structure which is an inevitable consequence of the requirement of a diffeomorphism invariant field theory. 

This first example for an application of the possifold program already demonstrates its power and justifies its consideration as the tool we were seeking for at the beginning. However, this tool reveals the origin of further results. By a similar reasoning as for the derivation of the gravitational entropy bound, a generalized Ryu-Takayanagi prescription for entanglement entropies in arbitrary diffeomorphism invariant field theories was shown in \cite{Averin:2025zua}. We already mentioned this result in chapter \ref{Kapitel 1} and we will apply it in the next chapter as it reveals important insights concerning black hole physics and especially the information paradox. The important achievement the possifold program does here is to connect an entanglement entropy within a quantum theory $(\text{possibilities}, \text{ weight factors})$ with the specific subset of the set $(\text{possibilities}, \text{ weight factors})$ (,i.e., concretely the subregion of the theory's phase space) which is responsible for the particular entanglement entropy. The reasoning to understand the generalized Ryu-Takayanagi prescription in \cite{Averin:2025zua} follows hereby again the logic of the possifold program. The entanglement entropy is expressed as a sum of possibilities \eqref{1} (a functional integral over paths in phase space), is then reorganized into suited bundles which then reveal the Ryu-Takayanagi prescription as well as the relevant subregion in phase space responsible for the entanglement. This example demonstrates further the utility and power of the possifold approach. We will use the generalized Ryu-Takayanagi prescription as mentioned in the next chapter in order to find the relevant region in phase space responsible for black hole entropy. However, at this stage we are already able to emphasize an important point which has caused a large amount of confusion in the literature. We feel that the discussion of this chapter is helpful in resolving this confusion as it makes the sources of confusion manifest. The confusion is related to our discussion about not allowed questions in quantum theory which naively seem meaningful as they make sense in the theory's constructed illusion. We have already discussed a simple example. However, the following example will be less trivial. In the quantum mechanical sum \eqref{1}, we have to sum over the theory's set of possibilities, i.e. technically over the paths of the theory's phase space. Concretely, this means we are summing over a set of canonical variables. Phase space paths satisfying the equations of motion hereby can contribute dominantly to this sum and lead to the constructed illusion as if the system evolves along those trajectories. For gravitational theories, the constructed illusions are interpreted as spacetimes solving the gravitational field equations. Such spacetimes may for instance be black hole solutions. From this illusion of a spacetime, one may then be naively tempted to ask questions which, as we emphasized however, need not to make sense. We name a few examples of such questions. What was before the beginning of time in a Friedmann universe? What is in the interior of a black hole? What are its degrees of freedom and what is in the black hole singularity? All of these are considered deep and unsolved questions and we do not say here that they actually make no sense at all. We however point out that some of them may actually make no sense. To give a more concrete but simpler example, we can consider a Schwarzschild black hole in Einstein gravity. In Euclidean signature it is given by a cigar solution capturing the outside of the black hole. Hence, the region outside of the black hole is sufficient in describing its evolution as a path in Hamiltonian phase space. Put it differently, the black hole interior does not lead to independent canonical coordinates in a Hamiltonian description. The black hole interior is rather an artifact of the spacetime interpretation. It does not possess independent degrees of freedom one could ask for to be measured. It does not correspond to independent canonical coordinates to be summed over in the quantum mechanical sum \eqref{1}. Pretending naively the opposite does yield to contradictions and paradoxes. 

Indeed, in \cite{Almheiri:2020cfm} it was explained how the firewall paradox is an artifact of pretending the degrees of freedom of Hawking radiation escaping an evaporating black hole to be independent of the degrees of freedom of the black hole interior (see \cite{Almheiri:2012rt,Almheiri:2013hfa} for the original references on the firewall paradox as well as \cite{Almheiri:2020cfm} for its resolution). In this context, the region of the black hole interior which does not possess degrees of freedom independent of those of the escaped radiation is termed an island. We have here nothing new to say about the resolution of the firewall paradox as discussed in \cite{Almheiri:2020cfm}. What we want to point out is that the island concept appears within our discussion quite naturally (while it appears in the language of \cite{Almheiri:2020cfm} surprisingly which is due to referring only to spacetime but not to phase space). In terms of the Hamiltonian phase space the meaning of an island is a region of a spacetime constructed from the canonical coordinates which does not correspond to independent canonical coordinates. Naively pretending the opposite leads to misconceptions such as the firewall paradox or the wrong Page curve as reviewed in \cite{Almheiri:2020cfm}. Note that this is in line with our discussion here. A naive interpretation of the spacetime concept may be misleading. Instead, a proper understanding of the theory's phase space is necessary as it is the central object appearing in the quantum mechanical sum \eqref{1}. The possifold program is hereby helpful to precisely understand the phase space structure as we have seen in the examples above.
\\

We end this chapter by giving an outlook on further applications of the possifold program we are going to analyze somewhere else. We argue why we expect the possifold notion to be useful there.

In the classical limit, i.e. for $\hbar \to 0,$ every point in Hamiltonian phase space corresponds to a physically distinguishable state. It can be uniquely distinguished by measuring, for instance, its canonical coordinates, i.e. canonical positions and momenta. For finite $\hbar$ this is no longer true. This reflects basically Heisenberg's uncertainty principle. Phase space volumes can therefore be thought of as a measure of the number of quantum mechanically distinguishable states. As we have explained, for diffeomorphism invariant field theories the possifold program reveals severe bounds on certain phase space volumes. These bounds on phase space volumes formalize the gravitational entropy bound on the quantum mechanically distinguishable states inside a region within quantum gravity which is heuristically expected from black hole thermodynamics. While an understanding of the gravitational entropy bound utilizing the possifold notion is already devoted in \cite{Averin:2024yeo}, we note that there are further phenomena in physics concerning the limited quantum mechanical distinguishability of physical states compared to phase space. In quantum theories of non-Abelian gauge theories, one expects the phenomenon of confinement as well as the existence of a mass gap. An explanation whether and as to why this is the case is still an open problem in the literature. Indeed, the strong force in the standard model is described by quantum chromodynamics. While this theory's phase space contains colored states, colored states are not observed at low energies. Rather, quantum mechanically distinguishable states are such that colored constituents are confined to hadrons. This color confinement is a characteristic observed phenomenon of the strong interaction although it lacks a satisfactory theoretical description so far. The analogy with the gravitational entropy bound suggests that non-Abelian gauge theories also possess a phase space geometry which the possifold program could be able to reveal and which would then reflect the mentioned phenomena such as confinement or existence of a mass gap. Indeed, the AdS/CFT correspondence suggests that the gravitational entropy bound in the gravitational bulk theory is dual to confinement in the boundary gauge theory. Since the possifold notion is used in the proof of the gravitational entropy bound, this duality provides further evidence the possifold notion should also be central in explaining confinement in gauge theories.

Further questions where the possifold program might be enlightening are related to the famous problem of UV divergences in quantum theories of gravity. So far, we have defined a quantum theory as a sum over paths in its Hamiltonian phase space as required by \eqref{1}. Straightforwardly, we have also insisted that gravitational theories should be quantized the same way. For diffeomorphism invariant field theories, a careful manipulation of the so obtained phase space path integrals then indeed leads to the gravitational entropy bound \cite{Averin:2024yeo} and the generalized Ryu-Takayanagi formula \cite{Averin:2025zua}, i.e. statements a quantum theory of gravity is expected to obey. However, we have not shown so far that these quantum theories of gravity after all do make sense, i.e. whether they predict via \eqref{1} finite results for measurable quantities. Attempts to quantize gravity (see \cite{Kiritsis:2019npv} for a review and original references) show general relativity to be non-renormalizable. That is, quantizing Einstein gravity using a standard Feynman diagram summation leads to observables containing UV divergences. The appearance of UV divergences would imply measurable quantities to be very sensitive to UV length scales. 

Obviously, since gravitational phenomena are experienced to be finite in Nature, such a high UV sensitivity is not observed. Then there arises the question as to what tames this UV sensitivity in gravity? We have already noted in the last chapter that such seeming sensitivities on length scales could be an artifact of an improperly organized summation. The seeming UV sensitivity is indeed concluded by organizing perturbative expansions according to Feynman diagrams. The spacetime notion of Feynman diagrams does however not explicitly capture central gravitational properties such as black hole formation. Although not directly apparent in the perturbative expansion, they might tame the UV sensitivity and measurable quantities could then be free of UV divergences. In such a case, the Feynman diagram expansion should be reorganized to make this taming evident at the possible cost of not relying apparently on the spacetime notion. As we argued in the last chapter, the proper reorganization of the quantum mechanical sum \eqref{1} is precisely the goal of the possifold program. Indeed, utilizing the possifold notion directly leads to the gravitational entropy bound for diffeomorphism invariant field theories which bounds the phase space volume and hence the contribution to phase space path integrals of certain regions in phase space. 

For Einstein gravity with minimally coupled matter the gravitational entropy bound shows \cite{Averin:2024yeo} that in this way the contribution of short length scale processes to the quantum mechanical sum of possibilities \eqref{1} is relatively suppressed compared to large length scale processes. This might explain why measurable quantities are less sensitive on short length scales than one would naively expect from a conventional field theory counting.

Despite UV divergences in attempts to quantize gravity, there are more examples where a Feynman diagram expansion leads to a high sensitivity on short length scales. This occurs so for the mass of the Higgs boson in the standard model or for the cosmological constant. However, in all of these cases this prediction is not what is observed in Nature. Without a fine-tuning of parameters, the high sensitivity on short length scales leads to the expectation of a large value for the Higgs mass (whose failure with observations is known as the gauge hierarchy problem) as well as the cosmological constant (whose failure with observations is known as the cosmological constant problem). We note that UV divergences in quantization of gravity, the hierarchy problem as well as the cosmological constant problem all share the same mathematical origin. The sensitivity of observables on short length scales suggested by a perturbative expansion with Feynman diagrams is higher than observed. This hence leads to the expectation that there is a so far missed mechanism which is able to tame the high UV sensitivity and which applies in all of these problems. 

In our discussion here, we have already encountered what this mechanism might be. Indeed, the gravitational entropy bound provides a taming on UV sensitivity for Einstein gravity with minimally coupled matter, i.e. in the validity of the equivalence principle, as explained above. This taming is also not directly evident in a perturbative Feynman diagram expansion. However, whether this taming is enough to address any of the mentioned problems remains to be further investigated.

The last problem we want to mention is that of black hole microstates. What are the states responsible for the black hole entropy? When asking this important question, we should clarify what answer we can expect. We cannot hope for a description of those states similar to, for instance, the eigenstates of the hydrogen atom, i.e. analytic expressions for superpositions of position eigenstates. Such an analytic treatment is barely possible in single-particle quantum mechanics not to speak of quantum field theory. Instead, we have already explained what we do aim for. We want to concretely list the subregion of the phase space of a gravitational theory which is responsible for black hole microstates, i.e. which region is relevant in the phase space path integration when computing observables related to black hole microstates? We already discussed in detail that the possifold notion is central in the derivation of the generalized Ryu-Takayanagi formula which is in turn a central tool for obtaining this phase space region. While we perform the application of the generalized Ryu-Takayanagi formula to this problem in the next chapter, a complete detailed listing of the relevant phase space region will be investigated somewhere else.

As we will see in the next chapter, such a listing can reveal black hole excitations which might be hidden in a naive spacetime description. While the implications of such excitations are interesting for black holes, they are also important for analogous spacetimes such as de Sitter spacetime or our universe. However, before we can turn to cosmological implications, we first focus on a treatment of the black hole case. We begin our treatment in the next chapter.   

\section{Microscopic Origin of Page Curve}
\label{Kapitel 3}

In this chapter, we will apply the generalized Ryu-Takayanagi formula \cite{Averin:2025zua} to black holes in order to infer what degrees of freedom are responsible for their entropies. The generalized Ryu-Takayanagi formula is able to do so as it makes direct contact with the phase space of the considered theory. As we have emphasized, it is the possifold notion which is able to reveal this contact. With the application of the generalized Ryu-Takayanagi formula in this chapter, we want to demonstrate two things.

First, it provides an example of an application of the generalized Ryu-Takayanagi formula shown in \cite{Averin:2025zua} to a situation beyond the context of the AdS/CFT correspondence to which the standard Ryu-Takayanagi type prescriptions are limited. 

Second, the application of the generalized Ryu-Takayanagi formula to black holes reveals what excitations are responsible for entropy and black hole microstates which is the core of the black hole information paradox. Hence, the possifold notion, here in the form of the generalized Ryu-Takayanagi formula, unveils black hole excitations which might be hidden in a naive spacetime interpretation. 

The information paradox can thus be seen as an artifact of a naive spacetime interpretation which is made evident by the possifold notion. This is precisely in line with the type of failures necessitating the use of the possifold notion as we have discussed in the overview provided in the preceding chapters. We hence devote this last chapter a further demonstration of the power of the possifold notion when applied to the information paradox. This is our main novel result here. 

While our overview has been entirely conceptual so far, the application of the generalized Ryu-Takayanagi formula requires an employment of the full formalism for a moment. However, the technical derivation will be rather short but the conclusions will insightfully fit our discussion. They reveal the misconceptions leading to the information paradox and how the possifold notion makes these misconceptions evident. A reader following the conceptual overview provided here without having the formal background of \cite{Averin:2024its,Averin:2024yeo,Averin:2025zua} may skip the short derivation in a first reading and move on to the conclusions. For an understanding how the conclusions are reached, the formalism and terminology of \cite{Averin:2024its} is necessary to which the reader is referred. Furthermore, the reader should be aquainted with the use of the generalized Ryu-Takayanagi prescription. What the generalized Ryu-Takayanagi prescription is computing and how it is used is explained in detail in \cite{Averin:2025zua}. To follow the entire proof of the generalized Ryu-Takayanagi prescription, the reader is required to have mastered all of \cite{Averin:2025zua} together with the relevant parts of \cite{Averin:2024yeo} indicated in \cite{Averin:2025zua}.

From now on, we will use for the moment of the application of the generalized Ryu-Takayanagi formula the formalism, terminology and notation of \cite{Averin:2024its,Averin:2024yeo,Averin:2025zua}. That is, we assume the reader to be familiar with the references at least in the manner described in the last paragraph. We consider a diffeomorphism invariant field theory $(\Gamma, \Theta, H)$ on a spacetime $M = \mathbb{R} \times \Sigma.$ Suppose, it admits a stationary black hole solution with bifurcate Killing horizon. The Hamiltonian $H$ of the theory describes the time evolution of the canonical coordinates on the hypersurface $\Sigma$ foliating the spacetime $M.$ We can choose the foliation such that for the particular black hole solution the hypersurface $\Sigma$ crosses the bifurcation surface at a fixed instant of time. At this moment of time, the black hole solution is then described by a state in phase space on $\Sigma$ which we denote as $\Phi \in \Gamma.$ Choices of codimension-2 surfaces $B \subseteq \Gamma$ divide $\Sigma$ and we are interested in computing the associated entanglement entropies. They are given by the generalized Ryu-Takayanagi formula. To leading order in $\hbar,$ we can infer the entanglement entropies using the prescription in chapter $3.1$ summarized by the equation $(33)$ in \cite{Averin:2025zua}. If $B \subseteq \Sigma$ is chosen to be the bifurcation surface, this yields the entanglement entropy 

\begin{equation}
\frac{1}{\hbar} K[\Phi^{(B)}].
\label{3}
\end{equation}

Hereby, $\Phi^{(B)} \in \Gamma^{(B)}$ denotes the action of the possifold flow $\partial \Sigma \to B$ on the state $\Phi \in \Gamma.$ That \eqref{3} fulfills the extremization condition of equation $(33)$ in \cite{Averin:2025zua} can be directly read off from the black hole spacetime solution represented by $\Phi \in \Gamma.$ Since the diffeomorphism generated by $K$ coincides near $B$ with the Killing vector field, the limit in $(33)$ in \cite{Averin:2025zua} can be ignored and the extremization procedure performed directly within the black hole spacetime solution.\footnote{As discussed in the proof of the generalized Ryu-Takayanagi formula, the choice of the canonical 1-form $\Theta$ defining the theory determines which surface deformations are allowed in the search for the extremal surface determining the entanglement entropy. We require here that $\Theta$ is chosen such that arbitrary deformations are allowed.}

The conclusions of determining this entanglement entropy are far reaching. We first note that the entanglement entropy of the black hole state across the bifurcation surface equals the black hole's Wald entropy \cite{Wald:1993nt,Iyer:1994ys}. Accordingly, the region of phase space responsible for this entanglement entropy and also for the black hole microstates must be contained in $\Gamma^{(B)}$ as is shown in the derivation of \eqref{3}. The states in this phase space region $\Gamma^{(B)}$ can be distinguished by their bifurcation surface charges as follows from the proof of Lemma $3.1.1$ in \cite{Averin:2024its}.

This is our main result here. Why is this important and what does it mean concretely? Consider two solutions of the theory's equations of motion. If they differ in their surface charges evaluated on $B,$ they correspond to different states in phase space. That is, they contribute generically as different possibilities in the quantum mechanical sum \eqref{1}. They generically lead to different quantum mechanical states to be distinguished by suited observables. And according to our main result, the states differing by their bifurcation surface charges must account for the Wald entropy of black holes in arbitrary diffeomorphism invariant field theories. For readers not being familiar with the concept of surface charges, a brief review is provided in \cite{Averin:2024its} and references therein. For our discussion here, however, it is sufficient to know that they are given by certain surface integrals over chosen codimension-2 surfaces with the integrand being determined by the theory's canonical 1-form $\Theta.$

Especially, we learn that there is no choice whether diffeomorphism invariant field theories contain the degrees of freedom necessary to account for the entropy of their black holes. They are not to be added by hand. Diffeomorphism invariance being the origin of the generalized Ryu-Takayanagi formula forces them to contain those black hole degrees of freedom in the mentioned way.

For black holes formed in gravitational collapse and undergoing subsequent evaporation by Hawking radiation, it is argued in \cite{Almheiri:2020cfm} that application of the quantum extremal surface prescription reveals a time evolution of entanglement entropy consistent with the Page curve and hence quantum mechanical unitarity (see \cite{Almheiri:2019psf,Penington:2019npb,Almheiri:2019hni,Penington:2019kki,Almheiri:2019qdq} for the original references). While the used prescription is able to deduce the decreasing part of the Page curve from the semiclassical geometry, it does not make a statement about the microscopic origin of the necessary black hole degrees of freedom responsible for the entanglement entropy. The point we want to add here to this discussion is that the generalized Ryu-Takayanagi formula by keeping the contact to the theory's phase space precisely fills this gap. It reveals the microscopic origin of those black hole degrees of freedom and they are for any diffeomorphism invariant field theory inevitably given in the form inferred above. 

Finding the black hole degrees of freedom responsible for microstates and entropy is at the core of the information paradox. Our discussion teaches us that the information paradox is not a problem about physical theories but about mathematical consistency. This means it is not a question about whether and how a gravitational theory should be completed as to account for black hole degrees of freedom to ensure preservation of information. As we have seen, all diffeomorphism invariant field theories necessarily do contain those degrees of freedom since this turns out to be a constraint enforced by diffeomorphism invariance. Rather, the remaining question is whether the states in phase space with different bifurcation surface charges are indeed accounting for the black hole entropy and hence represent the searched black hole degrees of freedom. In other words, this question corresponds to the mathematical problem of giving  an explicit list of the phase space states responsible for black hole microstates. According to our results, we can in principle get such a list by evaluating the bifurcation surface charges of solutions of equations of motion underlying the considered diffeomorphism invariant field theory. By consistency, the states so obtained must account for black hole microstates. As we have already mentioned in the last chapter, we will return to this problem in a different place. 

An important objection at this point might be how to reconcile our findings here with the phrase of black holes possessing no hair? More precisely, for Einstein gravity in four dimensions the uniqueness theorems stay that any asymptotically flat and stationary solution of the vacuum field equations is given by the Kerr spacetime up to diffeomorphisms. This seems to suggest that there are actually no states in phase space to account for black hole entropy and microstates. Precisely this viewpoint led to the conclusion that black holes destroy the information about their formation sourcing the information paradox. However, our findings here show where this conclusion fails. Different solutions of the field equations related by gauge transformations of the considered theory (such as diffeomorphisms) can possess different bifurcation surface charges. As such, they correspond to different states in phase space and can lead to different quantum mechanical states distinguishable by measurements of suited observables. The wrong conclusion that all diffeomorphisms correspond to gauge redundancies can again be seen as an artifact of a naive spacetime interpretation. In fact, a small subset of a theory's gauge transformations may lead to physical excitations, i.e. different states in phase space. The possifold notion reveals hereby in the form of our findings a practical tool to single out the relevant excitations. 

Whether the phase space states provided by black hole solutions of field equations related by gauge transformations (such as diffeomorphisms or local supersymmetry transformations in supergravity theories) with different bifurcation surface charges are really enough to account for black hole entropy deserves further investigation as we have already explained. However, earlier attempts show that in this way suited states are indeed obtained \cite{Averin:2018owq} and that those states might indeed reproduce black hole entropy \cite{Averin:2019zsi}. In fact, these approaches were the motivation for the development of the possifold notion in \cite{Averin:2024its}. Having again demonstrated its power in this chapter, we close our discussion here.

Since we have already given an outlook on further applications of the possifold notion to be investigated, we end here by summarizing it in its central appeal:
\\

Possibilities in the quantum mechanical sum, unite as possifolds, and reveal Nature's unity!

\section*{Acknowledgements}
We thank Alexander Gußmann for many discussions on this and other topics in physics.

\end{document}